\documentclass[a4paper,11pt]{article}
\usepackage{pos,bm}

\usepackage{rotating}
\usepackage{graphicx,hyperref}
\definecolor{darkgreen}{rgb}{0,0.6,0}
\definecolor{pink}{rgb}{1,0.6,1}
\definecolor{purple}{rgb}{0.9,0,0.9}
\newcommand{\pvec}{\bm{p}}
\newcommand{\beq}{\begin{equation}}
\newcommand{\eeq}{\end{equation}}
\newcommand{\beqs}{\begin{eqnarray}}
\newcommand{\eeqs}{\end{eqnarray}}
\newcommand{\bitem}{\begin{itemize}\item}
\newcommand{\eitem}{\end{itemize}}

\newcommand{\me}[3]{\langle #1\vert\ #2\ \vert #3\rangle}

\newcommand{\xvec}{\mathbf{x}}

\newcommand{\dvec}{\bm{d}}
\newcommand{\Pvec}{\bm{P}}

\title{Progress on Meson-Baryon Scattering}
\author*[a]{Colin Morningstar}
\author[b]{John Bulava}
\author[c]{Andrew D.\ Hanlon}
\author[d]{Ben H\"orz}
\author[e]{Daniel Mohler}
\author[f]{Amy Nicholson}
\author[a]{Sarah Skinner}
\author[d]{Andr\'e Walker-Loud}

\affiliation[a]{Department of Physics, Carnegie Mellon University,
  Pittsburgh, PA, 15213, USA}

\affiliation[b]{Deutsches Elektronen-Synchrotron (DESY), Platanenallee 6, 15738, Zeuthen, Germany}

\affiliation[c]{Physics Department, Brookhaven National Laboratory, Upton, New York, 11973, USA}

\affiliation[d]{Nuclear Science Division, Lawrence Berkeley National Laboratory,
  Berkeley, CA, USA 94720}

\affiliation[e]{Helmholtzzentrum f\"ur Schwerionenforschung (GSI),
   Planckstrasse 1, 64291 Darmstadt, Germany}

\affiliation[f]{Department of Physics and Astronomy, University of North Carolina,
   Chapel Hill, NC, 27599, USA}

\emailAdd{cmorning@andrew.cmu.edu}

\abstract{Progress in computing various meson-baryon scattering amplitudes is presented 
on a single ensemble from the Coordinated Lattice Simulations (CLS) consortium with 
$m_\pi=200$~MeV and $N_f=2+1$ dynamical fermions. The finite-volume L\"uscher approach 
is employed to determine the lowest few partial waves from ground- and excited-state 
energies computed from correlation matrices rotated in a single pivot using a generalized
eigenvector solution. This analysis requires evaluating matrices of correlation functions
between single- and two-hadron interpolating operators which are projected onto definite 
spatial momenta and finite-volume irreducible representations. The stochastic LapH method
is used to estimate all needed quark propagators. Preliminary 
results are presented for $I=\frac{1}{2}, \frac{3}{2}$ $N\pi$ amplitudes including the 
$\Delta(1232)$ resonance and the $I=0$ $S$-wave amplitude with unit strangeness 
relevant for the $\Lambda(1405)$.}

\FullConference{%
 The 38th International Symposium on Lattice Field Theory, LATTICE2021
  26th-30th July, 2021
  Zoom/Gather@Massachusetts Institute of Technology
}

\begin{document}
\maketitle

\section{Overview}
This talk is a progress report on our efforts to determine meson-baryon and baryon-baryon
scattering parameters in a large number of flavor channels. Preliminary results 
from correlation matrices obtained using only one time source are 
presented for $I=\frac{1}{2}, \frac{3}{2}$ $N\pi$ amplitudes including the 
$\Delta(1232)$ resonance and the $I=0$ $S$-wave amplitude with unit strangeness 
relevant for the $\Lambda(1405)$ using the D200 ensemble from the Coordinated Lattice 
Simulations (CLS) consortium with $m_\pi=200$~MeV and $N_f=2+1$ dynamical fermions. 
Results from correlators estimated using four time sources will soon be available.

Some motivations for this work are as follows.
Meson-baryon amplitudes are useful for a variety of phenomenological applications both
at the physical pion mass $m_{\pi}^{\rm phys}$ and for chiral effective field theories 
(EFT) at varying pion masses $m_{\pi}$.  The process $\Delta(1232)\rightarrow N\pi$ 
is sometimes used as a degree-of-freedom in some EFT's.  The scattering lengths 
$a_{N\pi}^{I=3/2}$ and $a_{N\pi}^{I=1/2}$ will impact the discrepancy 
between lattice QCD and phenomenology determinations for $\sigma_{\pi N}$,
which is relevant for dark matter direct detection. Lattice QCD is a good
laboratory to study the $\Lambda(1405)$ by varying quark masses.  

\section{Methodology}
The finite-volume L\"uscher approach\cite{Luscher:1990ux,Rummukainen:1995vs,
Kim:2005gf,Briceno:2014oea} is employed to determine the lowest few partial 
waves from ground- and excited-state energies computed from correlation matrices rotated 
in a single pivot using a generalized eigenvector solution.  
Our implementation of the L\"uscher method uses the ``box matrix'' 
$B$ introduced in Ref.~\cite{Morningstar:2017spu}, along with the scattering $K$-matrix,
to form the energy quantization condition.  Parameters in the $K$-matrix are determined
using the determinant residual method of Ref.~\cite{Morningstar:2017spu}.
This analysis requires evaluating matrices of correlation functions
between single- and two-hadron interpolating operators which are projected onto definite 
spatial momenta and finite-volume irreducible representations. The stochastic LapH 
method\cite{Morningstar:2011ka} is used to estimate all needed quark propagators. 

Finite-volume stationary-state energies are obtained from temporal correlations 
$C_{ij}(t)=\langle 0\vert \overline{O}_i(t) O_j(0)\vert 0\rangle$, where $O_j(t)$
are appropriate single- and multi-hadron operators.
In finite volume, such energies are discrete, so the correlators can be expressed
in terms of the energies using   
   \beq
   C_{ij}(t) = \sum_n Z_i^{(n)} Z_j^{(n)\ast}\ e^{-E_n t},
   \qquad\quad Z_j^{(n)}=  \me{0}{O_j}{n},
   \eeq
ignoring negligible effects from the temporal boundary. 
It is not practical to do fits using the above
form, so we define a new correlation matrix $\widetilde{C}(t)$ 
using a single-pivot rotation
   \beq
   \widetilde{C}(t) = U^\dagger\ C(\tau_0)^{-1/2}\ C(t)\ C(\tau_0)^{-1/2}\ U,
   \eeq
where the  columns of $U$ are the eigenvectors of
   $C(\tau_0)^{-1/2}\,C(\tau_D)\,C(\tau_0)^{-1/2}$.
We choose $\tau_0$ and $\tau_D$ large enough so that $\widetilde{C}(t)$ 
remains diagonal for $t>\tau_D$ and such that the extracted energies are
insensitive to increases in these parameters.
Two-exponential fits to the diagonal elements $\widetilde{C}_{\alpha\alpha}(t)$
yield the energies $E_\alpha$ and overlaps $Z_j^{(n)}$.  However, 
energy shifts from non-interacting levels can be more accurately obtained 
using single-exponential fits to suitable \textit{ratios} of correlators.

It is extremely important to use judiciously constructed operators $O_j(t)$.
Our operator construction is detailed in Refs.\cite{PhysRevD.72.094506,PhysRevD.88.014511}.
Individual hadron operators are constructed using basic building blocks which
are covariantly-displaced LapH-smeared quark fields.  Stout link 
smearing\cite{PhysRevD.69.054501} is used for the displacements, and
Laplacian-Heaviside (LapH) smearing is used for the quark fields:
   \beq
    \widetilde{\psi}_{a\alpha}(x) =
      {\cal S}_{ab}(x,y)\ \psi_{b\alpha}(y),
     \qquad {\cal S} = 
     \Theta\left(\sigma_s^2+\widetilde{\Delta}\right),
   \eeq
where the three-dimensional gauge-covariant Laplacian $\widetilde{\Delta}$ is
given in terms of the smeared link variables $\widetilde{U}$, and $\sigma_s$ is
a smearing cutoff which determines the number $N_{\rm ev}$ of LapH eigenvectors
to retain.  The quarks are combined into so-called elemental meson and baryon operators:
 \beqs
 \overline{\Phi}_{\alpha\beta}^{AB}(\pvec,t)&=&
 \textstyle\sum_{\bm{x}} e^{i\pvec\cdot(\xvec+\frac{1}{2}(\bm{d}_\alpha+\bm{d}_\beta))}
   \delta_{ab}\ \overline{q}^B_{b\beta}(\bm{x},t)\ q^A_{a\alpha}(\bm{x},t),
 \\
  \overline{\Phi}_{\alpha\beta\gamma}^{ABC}(\pvec,t)&=& 
 \textstyle\sum_{\bm{x}} e^{i\pvec\cdot\xvec}\varepsilon_{abc}
\ \overline{q}^C_{c\gamma}(\bm{x},t)
\ \overline{q}^B_{b\beta}(\bm{x},t)
\ \overline{q}^A_{a\alpha}(\bm{x},t),
\eeqs
then the hadron operators are superpositions of the elemental
operators obtained by group-theory projections onto the irreducible representations
(irreps) of the appropriate lattice symmetry group:
\beq
  \overline{M}_{l}(t)= c^{(l)\ast}_{
 \alpha\beta}\ \overline{\Phi}^{AB}_{\alpha\beta}(t)\qquad\qquad
  \overline{B}_{l}(t)= c^{(l)\ast}_{
 \alpha\beta\gamma}\ \overline{\Phi}^{ABC}_{\alpha\beta\gamma}(t).
\eeq
For an operator creating a definite momentum $\pvec$, our operators transform
according to irreps of the little group of $\pvec$ on a cubic lattice.
For multi-hadron operators, it is very important to use superpositions of products 
of single-hadron operators of definite momenta.  Spatially localized multi-hadron
operators produce signals with large amounts of excited-state contamination.
Our hadron operator construction is very efficient and generalizes to
three or more hadrons.  Note that to speed up our computations to achieve the
statistics needed for extracting the low-lying energies required for our 
baryon-baryon scattering studies, we have not included any single hadron 
operators with quarks that are displaced from one another.

Including multi-hadron operators in our correlation matrices requires the
use of time-slice to time-slice quark propagators.  To make the calculations
feasible, we resort to employing stochastic estimates of such quark propagators.
The stochastic LapH method\cite{Morningstar:2011ka} is used.
We introduce $N_R$ vectors of $Z_4$ noise $\eta^{(r)}$ in the LapH subspace
   \beq
        \eta^{(r)}_{\alpha k}(t),\qquad t=\mbox{time},
         \ \alpha=\mbox{spin},\ k=\mbox{eigenvector number}.
   \eeq
We carry out variance reduction using noise dilution, which introduces
projectors $P^{(a)}$.  Defining
  \beq \eta^{[a]}=P^{(a)}\eta,\qquad X^{[a]}=D^{-1}\eta^{[a]},
  \eeq
we obtain Monte Carlo estimates of the quark propagators via
  \beq
     D_{ij}^{-1}\approx \frac{1}{N_R}\sum_{r=1}^{N_R}\sum_a
     X^{(r)[a]}_i \eta_j^{(r)[a]\ast}.
  \eeq
We define four dilution schemes:
   \beq
   \begin{array}{lll}
        P_{ij}^{(a)} = \delta_{ij}, & a=0, & \mbox{(none)},\\
        P_{ij}^{(a)} = \delta_{ij}\delta_{ai}, & a=0,1,\dots,N\!-\!1, & \mbox{(full),}\\
        P_{ij}^{(a)} = \delta_{ij}\delta_{a, Ki/N}, & a=0,1,\dots,K\!-\!1, & \mbox{(interlace-$K$)},\\
        P_{ij}^{(a)} = \delta_{ij}\delta_{a, i\,{\rm mod}\, k}, & a=0,1,\dots,K\!-\!1, 
                 & \mbox{(block-$K$)}.
   \end{array}
  \eeq
We apply dilutions to the time indices (full for fixed sources, interlace for relative sources),
the spin indices (full), and the LapH eigenvector indices (interlace-16).

Our current computations make use of 2000 configurations of the CLS D200 ensemble, which 
employs a $64^3\times 128$ lattice with spacing $a\sim 0.065~\rm{fm}$ and open boundary 
conditions in time.  The quark masses are tuned such that $m_\pi\sim 200~\rm{MeV}$ and 
$m_K\sim 480~{\rm MeV}$.
For the LapH smearing, we use $N_{\rm ev}=448$.  We are currently extending our computations 
to include the following source times:
$t_0=35$ forward, $t_0=64$ forward and backward, and $t_0=92$ backward.
Our Wick contractions are efficiently performed using tensor contraction software which
exploits common subexpression elimination\cite{Horz:2019rrn,contraction_optimizer} and makes heavy use
of threaded batched BLAS routines.  The various isospin channels that we are computing
are listed in Table~\ref{tab:channels}, as well as the total numbers of correlators
that we are computing in each channel.

\begin{table}
\caption[TabOne]{The various isospin channels we plan to study using the CLS
D200 ensemble, and the number of correlators we will compute in each channel.
\label{tab:channels}}
\begin{center}
\begin{tabular}{lcr}\hline
   Isospin channel  & D200 Number of Correlators\\ \hline
%
%   D200 numbers
%
$I=0,\ S=0,\ NN$                                         &  8357   \\
$I=0,\ S=-1,\ \Lambda, N\overline{K}, \Sigma\pi$ (45 SH) &  8143   \\
$I=\frac{1}{2},\ S=0,$ $N\pi$                            &   696   \\
$I=\frac{1}{2},\ S=-1,$ $N\Lambda,N\Sigma$               & 17816   \\
$I=1,\ S=0,$ $NN$ (66 SH)                                &  7945   \\
$I=\frac{3}{2},\ S=0,$ $\Delta, N\pi$                    &  3218   \\
$I=\frac{3}{2},\ S=-1,$ $N\Sigma$                        & 23748   \\
$I=0,\ S=-2,$ $\Lambda\Lambda, N\Xi, \Sigma\Sigma$ (66 SH)&16086   \\
$I=2,\ S=-2,$ $\Sigma\Sigma$ (66 SH)                     &  4589   \\
Single hadrons (SH)                                      &    33  \\ \hline
\end{tabular}
\end{center}
\end{table}

Scattering parameters are extracted from finite-volume energies using
our implementation\cite{Morningstar:2017spu} of the L\"uscher method.
We parametrize the inverse of the $K$-matrix,  then determine
best-fit values of the parameters using the determinant residual method\cite{Morningstar:2017spu}
in which we minimize
\beq
   \Omega(\mu,A)\equiv \frac{\det(A)}{\det[(\mu^2+AA^\dagger)^{1/2}]},
\eeq
with $A=1-\widetilde{K}^{-1}B^{-1}$, where $\widetilde{K}$ is the $K$-matrix 
with threshold factors removed, and $B$ is the so-called box matrix.  We
typically use $\mu=1$.

\begin{table}
\caption[TabTwo]{(Left) The $(2J,L)$ partial wave content of various blocks of 
  the matrix $1-\widetilde{K}^{-1}B^{-1}$ labelled by little group irrep $\Lambda(\dvec^2)$
  for $N\pi$ states of $I=\frac{1}{2},\frac{3}{2}$. Note that the integer $\dvec^2$ 
  refers to total momentum $\Pvec^2=(2\pi/L_{\rm lat})^2 \dvec^2$ for a lattice
  volume $L_{\rm lat}^3$.   (Right) The elements of the
  scattering $K$-matrix, denoted by $K^{(J)}_{L}$, which must be parametrized
  for $L\leq 2$. See Ref.~\cite{PhysRevD.88.014511} for a description of the irrep labels.
\label{tab:JLcontent}}
\begin{center}
\begin{tabular}{ll}\hline
   $\Lambda(\dvec^2)$  & $(2J,L)$ content for $L\leq 2$\\ \hline
$H_g(0)$     &  $(3, 1)$\\
$H_u(0)$     &  $(3, 2),\ (5, 2)$\\
$G_{1g}(0)$  &  $(1, 1)$\\
$G_{1u}(0)$  &  $(1, 0)$\\
$G_{2g}(0)$  &     \\
$G_{2u}(0)$  &  $(5, 2)$\\
$G_1(1),\ G_1(4)$     &  $(1, 0),\ (1, 1),\ (3, 1),\ (3, 2),\ (5, 2)$\\
$G_2(1),\ G_2(4)$     &  $(3, 1),\ (3, 2),\ (5, 2)$\\
$G(2)$       &  $(1, 0),\ (1, 1),\ (3, 1),\ (3, 2),\ (5, 2)$\\
$F_1(3)$     &  $(3, 1),\ (3, 2),\ (5, 2)$\\
$F_2(3)$     &  $(3, 1),\ (3, 2),\ (5, 2)$\\
$G(3)$       &  $(1, 0),\ (1, 1),\ (3, 1),\ (3, 2),\ (5, 2)$\\ \hline
\end{tabular}
\hspace*{5mm}
\begin{tabular}{c@{\hspace{4mm}}c}\hline\\[-3mm]
  $J$ & $K^{(J)}_{L}$ needed for $L\leq 2$\\[2mm] \hline\\[-2mm]
  $\frac{1}{2}$ & $K^{(1/2)}_{0},\  K^{(1/2)}_{1}$\\[2mm]
  $\frac{3}{2}$ & $K^{(3/2)}_{1},\  K^{(3/2)}_{2}$\\[2mm]
  $\frac{5}{2}$ & $K^{(5/2)}_{2}$\\[2mm]\hline
\end{tabular}
\end{center}
\end{table}

In this talk, we present preliminary results for $N\pi$ scattering in the 
isoquartet and isodoublet nonstrange channels using only one time source.  
Recall that the $K$-matrix has the form $K^{(J)}_{L'S'a';LSa}$.  For one
channel, $a=a'=0$, and for $N\pi$, we have total spin $S=S'=\frac{1}{2}$.
Invariance under parity requires $(-1)^{L+L'}=1$ for $N\pi$, which is tantamount
to $L=L'$.  Hence, we can use the simplified notation $K^{(J)}_L$.  Restricting
to $L\leq 2$, the $(2J,L)$ partial wave content of various blocks of the matrix 
$1-\widetilde{K}^{-1}B^{-1}$ are listed in Table~\ref{tab:JLcontent}. The 
elements of the scattering $K$-matrix which must be parametrized for $L\leq 2$
are also listed in Table~\ref{tab:JLcontent}.

The finite-volume energies are determined from
ratio fits to single-pivot rotated correlators. Determinations of energy shifts 
$\Delta_E$ from non-interacting energies were checked for stability against 
variations of the single-pivot times $(\tau_0,\tau_D)$ and on increasing
the number of operators $n_{\rm ops}$.  We parametrized resonant amplitudes with 
a Breit-Wigner form and used leading-order effective range expansions for 
non-resonant amplitudes, that is, parametrizing them with a constant.
Covariance matrices and all statistical errors were estimated using bootstrap resampling
with $N_{B} = 800$ samples.  All elastic levels are included. Any level within 
$1\sigma$ of an inelastic threshold was not included.  Parametrizations were
provided for all $S$- and $P$-waves. Higher partial waves are ignored for now,
but will be included in our final analyses.  For the $\Lambda(1405)$ channel, 
we consider coupled channels with $\Sigma\pi,\ N\overline{K}$.
Mixing of two-hadrons with stable hadrons is included in the $I=\frac{1}{2}$ and 
$I=0$, $S=1$ channels.  The relevant stable hadron for $I=\frac{1}{2}$ is the nucleon, 
and for $I=0,\ S=1$, it is the $\Lambda(1115)$.

\begin{figure}
\begin{center}
 \includegraphics[width=\textwidth]{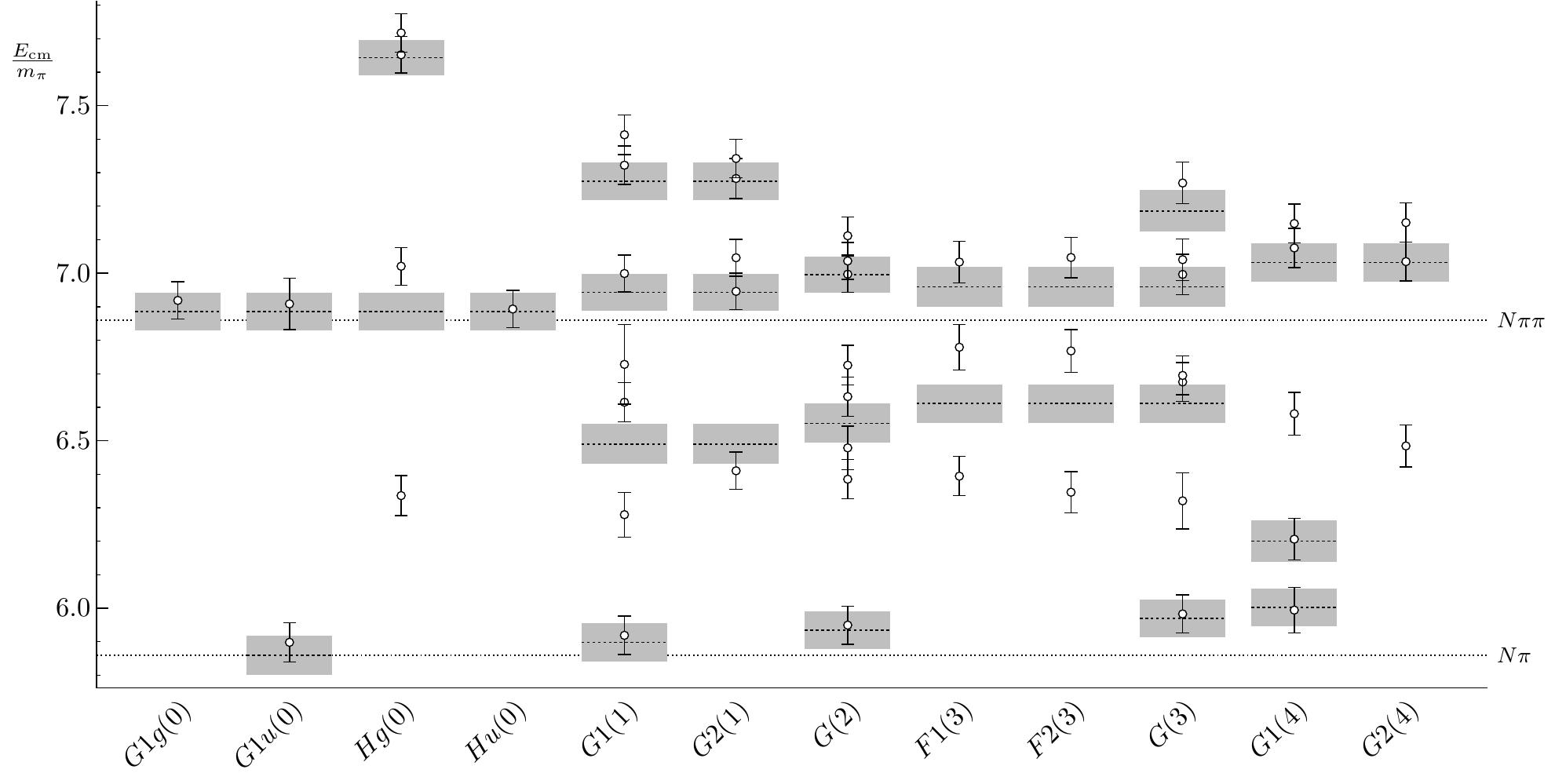}
\end{center}
\caption[FigOne]{Center-of-mass energies $E_{\rm cm}$ as
 ratios over the pion mass $m_\pi$ in the isoquartet non-strange sector for various little 
 group irreps.  The dashed horizontal lines show the non-interacting energies of
 the expected free two-particle states; the errors in the non-interacting energies
 are indicated by the gray boxes.  The integers in parentheses in the irreps
 indicate $\dvec^2$ for total momentum squared $\Pvec^2=(2\pi/L)^2 \dvec^2$.
 \label{fig:isoquartetspec}}
\end{figure}

\begin{figure}
  \includegraphics[width=0.52\textwidth]{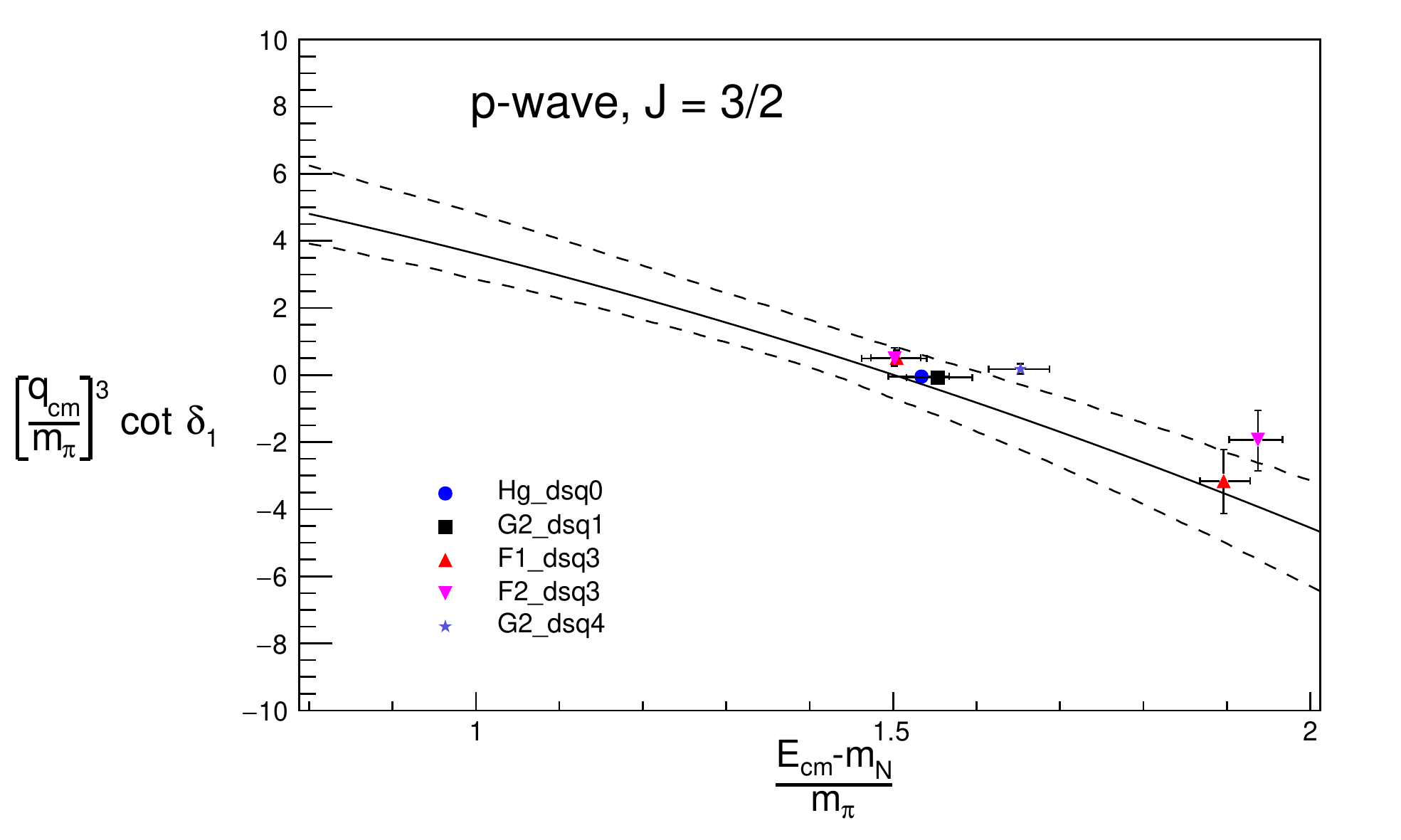}
  \includegraphics[width=0.46\textwidth]{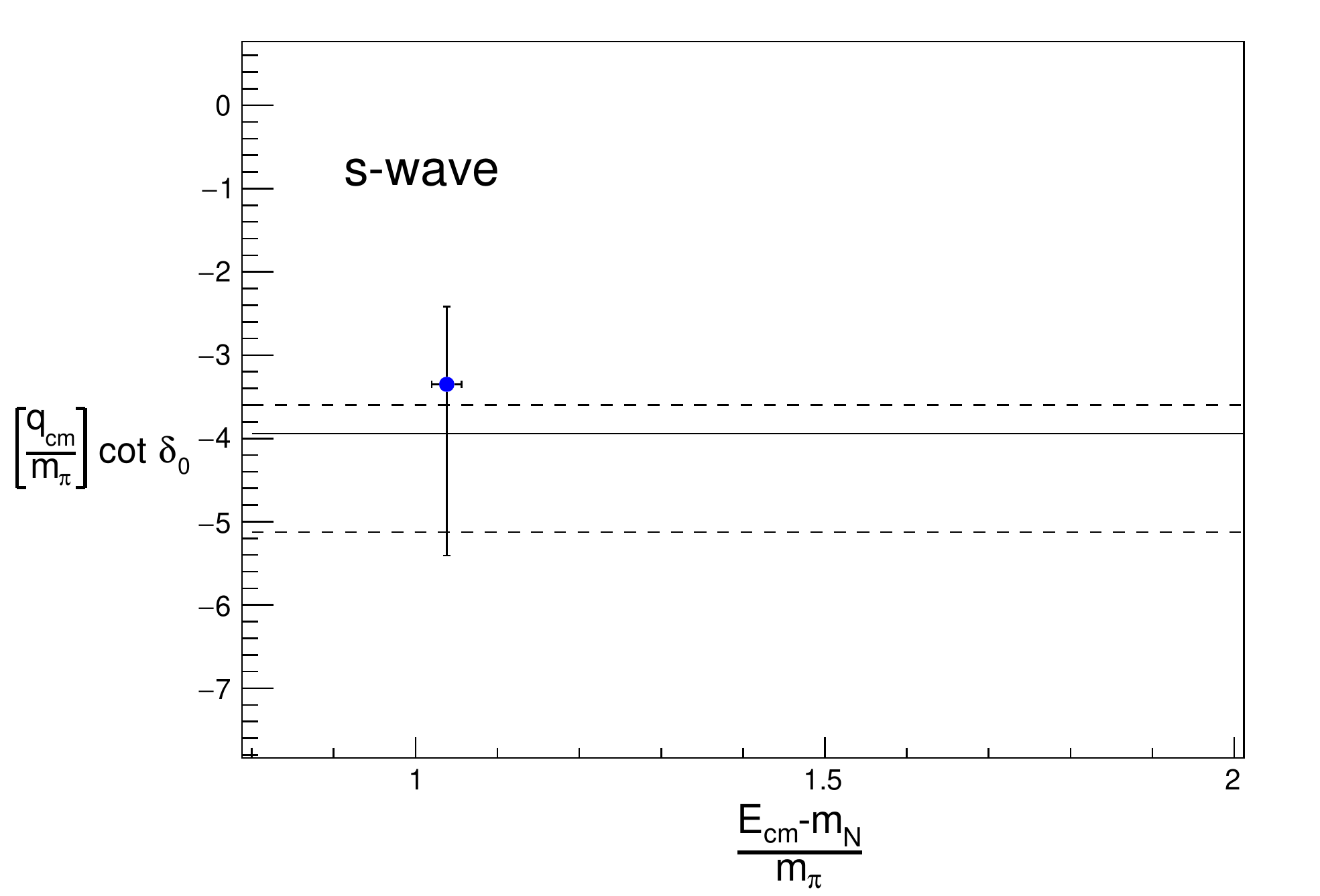}
\caption[FigTwo]{Threshold factors times cotangents of the phase shifts for
 the $P$-wave (left) and $S$-wave (right) for the isoquartet nonstrange channel
 against center-of-mass energies $E_{\rm cm}$ minus the nucleon mass $M_N$ as a ratio 
 over the pion mass $m_\pi$. Best-fit functions are shown as solid lines with error
 bands shown as dashed lines.
 \label{fig:phaseA}}
\end{figure}

The finite-volume spectrum for $I=\frac{3}{2}$ is shown in Fig.~\ref{fig:isoquartetspec}, 
and the scattering phase shifts are shown in Fig.~\ref{fig:phaseA}.  Seventeen levels
across $H_g(0)$, $G_{1u}(0)$, $G_1(1)$, $G(2)$, $F_{1}(3)$, $F_2(3)$, $G_1(4)$, $G_2(4)$
are included in the analysis.  The $G_{1g}(0)$ irrep which includes the leading 
$(2J,L)=(1, 1)$ 
wave is not included because the ground state in this irrep is inelastic. 
A Breit-Wigner form is used to parametrize
$\widetilde{K}_1^{(3/2)}(E)$, and constants are used for $\widetilde{K}_0^{(1/2)}(E)$ and 
$\widetilde{K}_{1}^{(1/2)}(E)$.  The best-fit results are
\begin{gather*}
    \frac{m_{\Delta}}{m_{\pi}} = 6.380(20), \quad g_{\Delta N\pi} = 13.7(1.5),
         \quad \chi^2/{\rm d.o.f.} = 1.74,
         \\ m_{\pi}a_0^{J=1/2} = -0.254(41), \quad (m_{\pi}a_1^{J=1/2})^{-1} = 2.61(44).
\end{gather*}
The resonance parameters are consistent with a fit to $P$-wave only irreps.

\begin{figure}
   \includegraphics[width=0.49\textwidth]{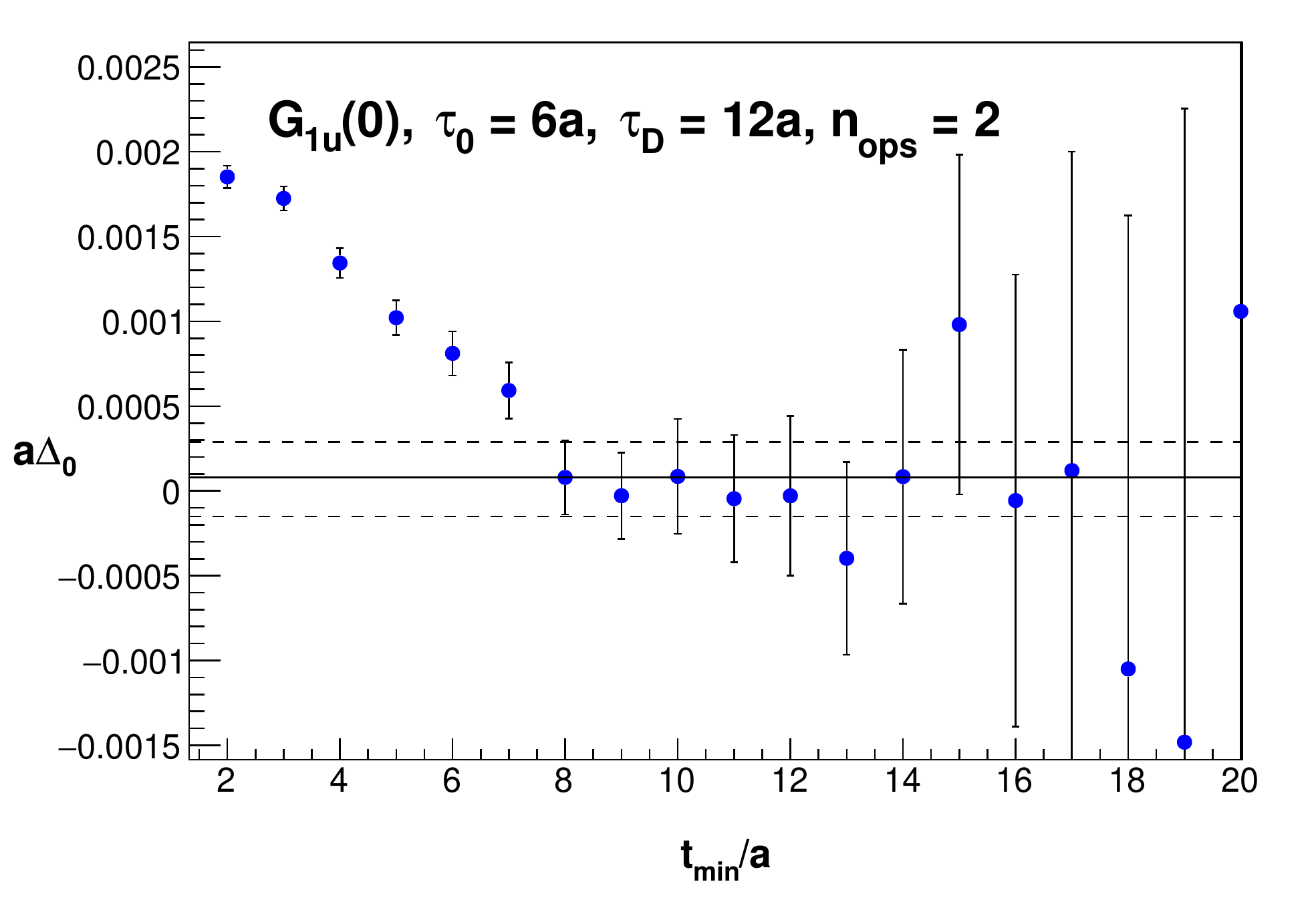}
   \includegraphics[width=0.49\textwidth]{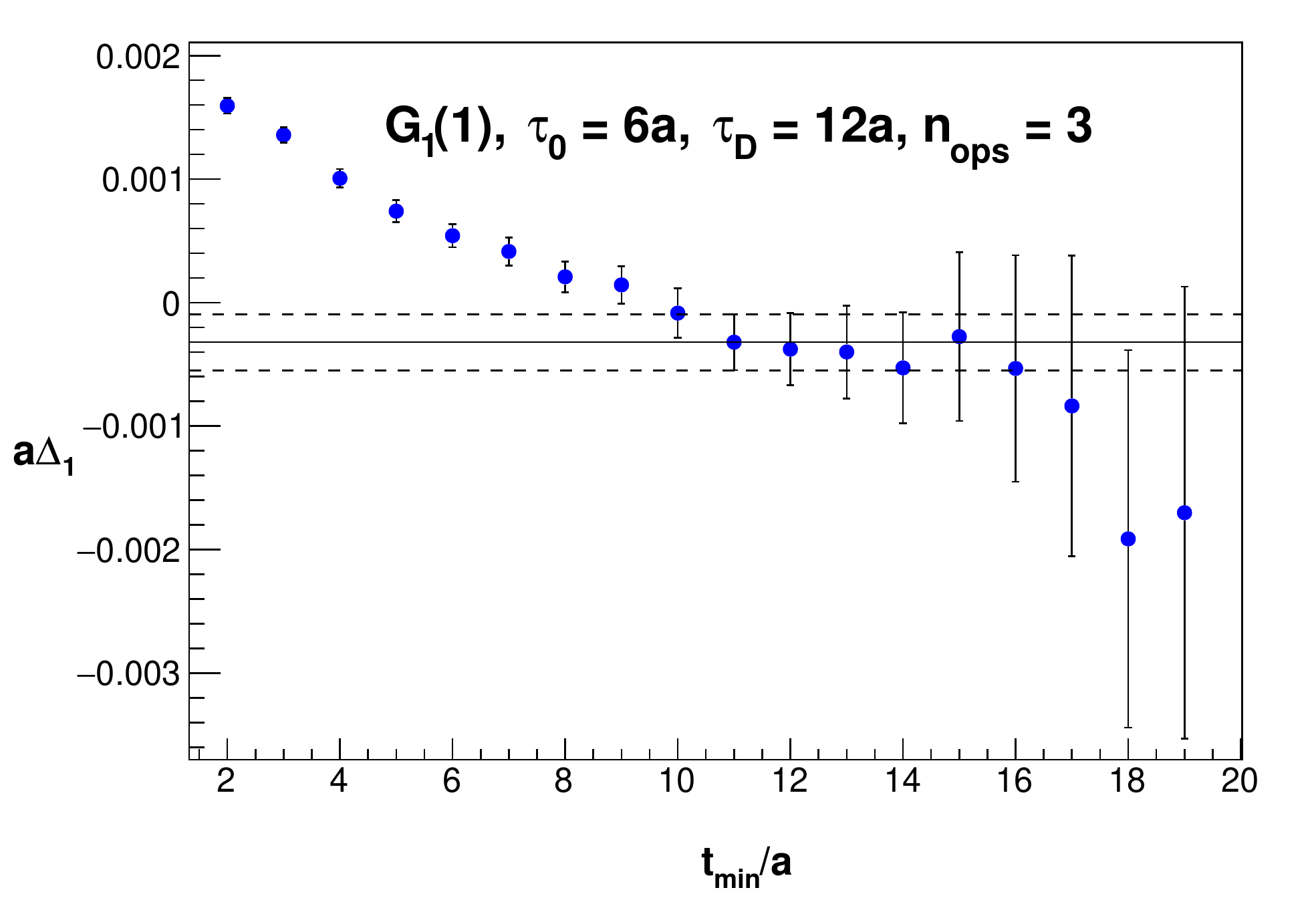}
\caption[FigThree]{
  $t_{\rm min}$ plots for the energy shifts from non-interacting two-particle energies in terms of 
  the lattice spacing $a$ for the isodoublet nonstrange channel for the ground state in 
  $G_{1u}(0)$ (left) and the first-excited state in $G_1(1)$ (right).  The single-pivot 
  rotation times $\tau_0,\tau_D$ are indicated, and $n_{\rm ops}$ are the numbers of 
  operators used in the correlation matrices.  Fits using $t_{\rm min}<8$ have
  $\chi^2/\rm{dof}>2$.
 \label{fig:isodoubleshifts}}
\end{figure}

Two examples of extracting energy shifts in the isodoublet nonstrange channel
are illustrated in the $t_{\rm min}$ plots shown in Fig.~\ref{fig:isodoubleshifts}.
Each point is a fit using a single-exponential form to the single-pivot rotated
correlator divided by the non-interacting level for time range from the $t_{\rm min}$ 
shown on the horizontal axes to $t_{\rm max}=25a$. Compared with the corresponding levels 
in the $I=\frac{3}{2}$ channel, the energy shifts are considerably smaller, as 
expected from the phenomenologically smaller value of the scattering length. 
A preliminary estimate of the $I=\frac{1}{2}$ scattering length $a_0^{1/2}$ is obtained 
by ignoring $L\ge 1$ contributions in the quantization conditions for both levels,
\begin{eqnarray}
     (m_\pi a_0^{1/2})^{-1} &=& -86.06(72.14),\qquad G_{1u}(0),\\
     (m_\pi a_0^{1/2})^{-1} &=& 19.82(16.25),\qquad G_{1}(1).
\end{eqnarray}
Additional statistics and a more complete 
analysis of the elastic spectrum for $I=\frac{1}{2}$ is required.   

\begin{figure}
\begin{center}
\includegraphics[width=0.4\textwidth]{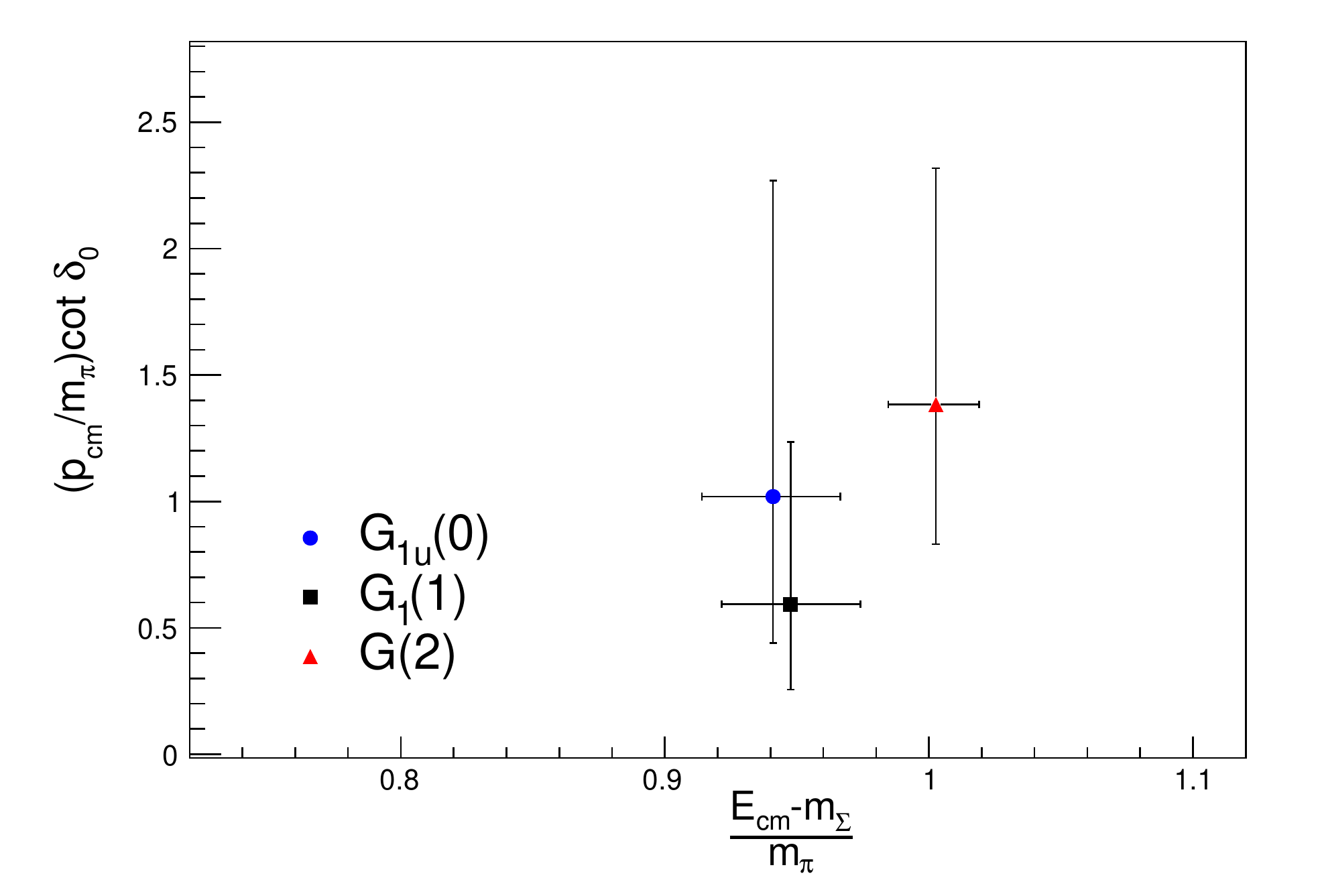}
\end{center}
\caption{\label{f:lambda} Preliminary estimates of the $(2J,L) = (1,0)$ $S$-wave 
scattering amplitude in the $I=0$, $S=1$ (Lambda) channel. These results are 
obtained from the leading-partial wave approximate in the $G_{1u}(0)$, $G_1(1)$ 
and $G(2)$ irreps using only the lowest scattering state, as discussed in the text.} 
\end{figure}

The final channel presented here is the $I=0,\ S=1$ channel, which is relevant for 
studying the $\Lambda(1405)$. This channel presents several additional difficulties. 
First, it may not be appropriate to truncate the quantization conditions at 
$L_{\rm max}=1$ due to the low-lying $\Lambda^*(1520)$ resonance in the $(2J,L)=(3,2)$ 
wave. Secondly, in order to accurately capture this excitation, it has been 
demonstrated~\cite{Meinel:2021mdj} that three-quark operators with 
gauge-covariant derivatives are needed to capture the orbital structure.
Such operators have not been included in our computations. 
Third, there are multiple coupled two-hadron scattering channels which are 
expected to mix significantly, the most important of which are $\Sigma\pi$ and 
$N\overline{K}$.  These issues complicate the construction of correlation matrices, 
the extraction of levels, and the parametrization of the coupled-channel scattering 
amplitude in this flavor channel.

All of these difficulties are ignored in this progress report which presents only our 
preliminary estimates.  Our results for the scattering amplitude using the $S$-wave
approximation are shown in Fig.~\ref{f:lambda}.  
Here, we assume that the lowest-lying $\Sigma\pi$ state is insensitive to the $(2J,L)=(3,2)$ 
wave in the $G_1(1)$ and $G(2)$ irreps, which also contain the $(1,0)$ and 
$(1,1)$ waves.  Only the lowest $\Sigma\pi$ scattering state is included in 
each of the $G_{1u}(0)$, $G_1(1)$, and $G(2)$ irreps.  As in the $I=\frac{1}{2}$ channel, the 
ground state in each irrep of non-zero total momenta is the lowest-lying stable $\Lambda$, which 
is far below the two-particle scattering threshold and so these levels are not included in 
the analysis.  
Despite the large statistical errors, there is some encouraging indication that the 
$\Sigma\pi$ phase shift is positive. 

For all of the results shown in this talk, we will soon present improved estimates using the 
increased statistics from including three more time sources and employing more 
comprehensive studies of the entire two-hadron spectra below the three-hadron thresholds, 
including all allowed types of two-hadron states.
\acknowledgments
Calculations for the results presented here were performed on the HPC
clusters ``HIMster II'' at the Helmholtz-Institut Mainz, ``Mogon II''
at JGU Mainz, and ``Frontera'' at the Texas Advanced Computing Center (TACC).  
The computations were performed using the \texttt{chroma\_laph} and 
\texttt{last\_laph} software suites.
\texttt{chroma\_laph} uses the USQCD \texttt{chroma}~\cite{Edwards:2004sx} library and 
the \texttt{QDP++} library.
The contractions were optimized with \texttt{contraction\_optimizer}~\cite{contraction_optimizer}.
The computations were managed with \texttt{METAQ}~\cite{Berkowitz:2017vcp,Berkowitz:2017xna}.
The correlation function analysis was performed with \texttt{chimera} and \texttt{SigMonD}.
We are grateful to our colleagues within the CLS initiative for sharing ensembles.

CJM acknowledges support from the U.S.~NSF under award PHY-1913158.
ADH is supported by the U.S. Department of Energy, Office of Science,
Office of Nuclear Physics through the Contract No. DE-SC0012704 and
within the framework of Scientific Discovery through Advance Computing
(SciDAC) award ``Computing the Properties of Matter with Leadership
Computing Resources''.  AWL acknowledges support from the U.S.~DOE through 
award number DE-AC02-05CH11231, an LBNL LDRD grant, and an Early Career 
Award. DM acknowledges funding by the Heisenberg Programme of the 
Deutsche Forschungsgemeinschaft (DFG, German Research Foundation), 
project number 454605793.

\bibliographystyle{JHEP}
\bibliography{references}

\providecommand{\href}[2]{#2}\begingroup\raggedright\begin{thebibliography}{10}

\bibitem{Luscher:1990ux}
M.~Luscher, \emph{{Two particle states on a torus and their relation to the
  scattering matrix}},
  \href{https://doi.org/10.1016/0550-3213(91)90366-6}{\emph{Nucl. Phys.}
  {\bfseries B354} (1991) 531}.

\bibitem{Rummukainen:1995vs}
K.~Rummukainen and S.A.~Gottlieb, \emph{{Resonance scattering phase shifts on a
  nonrest frame lattice}},
  \href{https://doi.org/10.1016/0550-3213(95)00313-H}{\emph{Nucl. Phys.}
  {\bfseries B450} (1995) 397}
  [\href{https://arxiv.org/abs/hep-lat/9503028}{{\ttfamily hep-lat/9503028}}].

\bibitem{Kim:2005gf}
C.H.~Kim, C.T.~Sachrajda and S.R.~Sharpe, \emph{{Finite-volume effects for
  two-hadron states in moving frames}},
  \href{https://doi.org/10.1016/j.nuclphysb.2005.08.029}{\emph{Nucl. Phys.}
  {\bfseries B727} (2005) 218}
  [\href{https://arxiv.org/abs/hep-lat/0507006}{{\ttfamily hep-lat/0507006}}].

\bibitem{Briceno:2014oea}
R.A.~Briceno, \emph{{Two-particle multichannel systems in a finite volume with
  arbitrary spin}},
  \href{https://doi.org/10.1103/PhysRevD.89.074507}{\emph{Phys. Rev.}
  {\bfseries D89} (2014) 074507}
  [\href{https://arxiv.org/abs/1401.3312}{{\ttfamily 1401.3312}}].

\bibitem{Morningstar:2017spu}
C.~Morningstar, J.~Bulava, B.~Singha, R.~Brett, J.~Fallica, A.~Hanlon et~al.,
  \emph{{Estimating the two-particle $K$-matrix for multiple partial waves and
  decay channels from finite-volume energies}},
  \href{https://doi.org/10.1016/j.nuclphysb.2017.09.014}{\emph{Nucl. Phys. B}
  {\bfseries 924} (2017) 477}
  [\href{https://arxiv.org/abs/1707.05817}{{\ttfamily 1707.05817}}].

\bibitem{Morningstar:2011ka}
C.~Morningstar, J.~Bulava, J.~Foley, K.J.~Juge, D.~Lenkner, M.~Peardon et~al.,
  \emph{{Improved stochastic estimation of quark propagation with Laplacian
  Heaviside smearing in lattice QCD}},
  \href{https://doi.org/10.1103/PhysRevD.83.114505}{\emph{Phys. Rev. D}
  {\bfseries 83} (2011) 114505}
  [\href{https://arxiv.org/abs/1104.3870}{{\ttfamily 1104.3870}}].

\bibitem{PhysRevD.72.094506}
S.~Basak, R.G.~Edwards, G.T.~Fleming, U.M.~Heller, C.~Morningstar, D.~Richards
  et~al., \emph{{Group-theoretical construction of extended baryon operators in
  lattice QCD}}, \href{https://doi.org/10.1103/PhysRevD.72.094506}{\emph{Phys.
  Rev. D} {\bfseries 72} (2005) 094506}.

\bibitem{PhysRevD.88.014511}
C.~Morningstar, J.~Bulava, B.~Fahy, J.~Foley, Y.C.~Jhang, K.J.~Juge et~al.,
  \emph{{Extended hadron and two-hadron operators of definite momentum for
  spectrum calculations in lattice QCD}},
  \href{https://doi.org/10.1103/PhysRevD.88.014511}{\emph{Phys. Rev. D}
  {\bfseries 88} (2013) 014511}.

\bibitem{PhysRevD.69.054501}
C.~Morningstar and M.~Peardon, \emph{{Analytic smearing of $\mathrm{SU}(3)$
  link variables in lattice QCD}},
  \href{https://doi.org/10.1103/PhysRevD.69.054501}{\emph{Phys. Rev. D}
  {\bfseries 69} (2004) 054501}.

\bibitem{Horz:2019rrn}
B.~H\"orz and A.~Hanlon, \emph{{Two- and three-pion finite-volume spectra at
  maximal isospin from lattice QCD}},
  \href{https://doi.org/10.1103/PhysRevLett.123.142002}{\emph{Phys. Rev. Lett.}
  {\bfseries 123} (2019) 142002}
  [\href{https://arxiv.org/abs/1905.04277}{{\ttfamily 1905.04277}}].

\bibitem{contraction_optimizer}
B.~H\"orz, ``Contraction optimizer.''
  \url{https://github.com/laphnn/contraction_optimizer}, 2019.

\bibitem{Meinel:2021mdj}
S.~Meinel and G.~Rendon, \emph{{$\Lambda_c \to \Lambda^*(1520)$ form factors
  from lattice QCD and improved analysis of the $\Lambda_b \to \Lambda^*(1520)$
  and $\Lambda_b \to \Lambda_c^*(2595,2625)$ form factors}},
  \href{https://arxiv.org/abs/2107.13140}{{\ttfamily 2107.13140}}.

\bibitem{Edwards:2004sx}
{\scshape SciDAC, LHPC, UKQCD} collaboration, \emph{{The Chroma software system
  for lattice QCD}},
  \href{https://doi.org/10.1016/j.nuclphysbps.2004.11.254}{\emph{Nucl. Phys. B
  Proc. Suppl.} {\bfseries 140} (2005) 832}
  [\href{https://arxiv.org/abs/hep-lat/0409003}{{\ttfamily hep-lat/0409003}}].

\bibitem{Berkowitz:2017vcp}
E.~Berkowitz, ``Metaq: Bundle supercomputing tasks.''
  \url{https://github.com/evanberkowitz/metaq}, 2018.

\bibitem{Berkowitz:2017xna}
E.~Berkowitz, G.R.~Jansen, K.~McElvain and A.~Walker-Loud, \emph{{Job
  Management and Task Bundling}},
  \href{https://doi.org/10.1051/epjconf/201817509007}{\emph{EPJ Web Conf.}
  {\bfseries 175} (2018) 09007}
  [\href{https://arxiv.org/abs/1710.01986}{{\ttfamily 1710.01986}}].

\end{thebibliography}\endgroup

\end{document}